\documentclass[a4paper]{article}

\newcommand{\sign}{\mathop{\rm sign}\nolimits}

\begin{document}
\title{Low energy physics \\ and properties of extra space}

\author{S.G. Rubin \\ National Research Nuclear University ''MEPhI'' ,  \\
E-mail: sergeirubin@list.ru}

\date{}
\maketitle

\begin{abstract}
The mechanism of low energy physics formation in the framework of multidimensional gravity is discussed. It is shown that a wide set of parameters of a primary theory could lead to the observable Universe.  Quantum fluctuations of extra space metric and its consequent classical evolution play an important role in this process.
\end{abstract}

\section{Introduction}

"A theory of everything" (TOE) is usually considered as a final goal. It means an existence of some universal lagrangian containing fields, symmetries and parameters (coupling constants) that should describe all phenomena in the Universe. But even if the goal was achieved main questions listed above remain unanswered:
\begin{itemize}
  \item What is the origin of the fields? Why just these fields are realized in the Nature?
  \item Why just these symmetries are realized but not others?
  \item Why the parameters have specific values?
  \item Why our space has specific number of dimensions (4, 10, 11, 26)?
  \item The situation is complicated by the known problem: the physical parameters acquire their observable values which are extremely fine tuned.
\end{itemize}
One may conclude that a TOE can not be a real final theory due to its internal reasons.

The alternative approach - multiverse - is also known. In this case it is supposed that many theories (universes) could be realized in Nature and we accidentally live in some space domain (Universe) described by one of the theory. This way is partly declared by the string theory. Main objection to this approach can be expressed shortly in one phrase: "If everything is possible, it is not a science". Let me suggest some example as a counterargument. Indeed, it would be strange if somebody  would explain the observed Earth mass just saying that a lot of planets with different masses do exist. Does it mean that any scientific researches in this direction are meaningless? Not at all. Instead we develop the theory of planet formation and some other scientific directions like the geology, the geophysics.

\section{Formation of low energy physics}

Our aim is to propose the approach that connects observable physical parameters and initial parameters of nonlinear gravity acting in $D-$ dimensional. It will be shown that a universe similar to ours can be formed for a wide range of initial parameters. In case of success, answers to the questions itemized above would be quite evident. The main tools are multidimensional gravity and quantum theory. The latter is needed to justify a quantum creation of manifolds with various metrics. For our purposes we need only the same fact of a nonzero probability of any metric formation in a small space region.

The essence of the approach is as follows. Suppose that a sufficient number of extra dimensions does exist. A metric of a whole space $M_0$ is chosen in the form
\begin{equation}\label{times0}
M_0 =M_4\times M.
\end{equation}
Here $M_4$ is our 4-dim space and $M$ is an extra space. It is supposed that every 4-dim lagrangian, e.g. the Standard Model lagrangian is deduced from a primary one that acts in the whole space $M_0$. Let this lagrangian contains initial parameters $a_{in}$. The observed values of the parameters $a_{obs}$ of the Standard Model are their functions
\begin{equation}\label{aobs}
a_{obs}=f(a_{in},g_M).
\end{equation}
Here $g_M$ is a metric tensor of the extra space.
A form of the function $f$ is a separate problem which was discussed partly in \cite{RuZin}. The internal metric $g_M$ is usually fixed by low energy physics what leads to strict connection between two sets $a_{in}$ and $a_{obs}$. Thus, in spite of many interesting applications have been performed, see e.g. \cite{Brokoru07, BOBRU}, the problem of origin of the same parameters $a_{in}$ remains. Let us make a next step and suppose that the geometry has the form
\begin{equation}\label{times1}
M_0 =M_4\times M\times K,
\end{equation}
where $K$ is an another extra space with a metric $g_K$. Formula (\ref{aobs}) is transformed into the two-step expression
\begin{equation}\label{aobs1}
a_{in}=f_1 (b_{in},g_K), \quad a_{obs}=f(a_{in},g_M),
\end{equation}
where $b_{in}$ is a new set of initial parameters. For the first sight nothing has been changed. In fact we have obtain additional freedom - an internal metric $g_K$ may be varied freely. The set $a_{in}$ can be obtained from a variety of sets $b_{in}$ by fitting the metric $g_K$. We will refer to the set $a_{in}$ as a secondary set and to the set $b_{in}$ as a primary one. Our nearest goal is to elaborate some production mechanism of various metrics $g_K$. It seems a trivial task because quantum fluctuations claimed produce any form of metric. The problem becomes much less simple if one takes into account a classical motion of an extra space metric $g_K$ just after its nucleation. As a result, a set of final metrics $g_K (t\rightarrow \infty)$ appears to be quite limited \cite{RuZin}. This strongly reduces a number of possibilities for the metric variation in (\ref{aobs1}).

As was shown in \cite{RuZin} the metrics could evolve classically to some stationary configuration $g_{stat}(y)=const$ that does not depend on an initial metric.  This way gives rise a large but limited number of an initial sets $a_{in}$ of parameters that lead to the universe similar to ours. The problem of fine tuning would be hardly solved.

%

Another wide class of theories could be obtained if classical equations for the metric $g_K$ have finite-dimensional attractors. In this case a behavior of the metric at large times depends on an arbitrary initial metric and form a continuous set. Thus we come to infinite set of effective theories.

So far we have implicitly assumed that the structure of the space is
$T\times M_{D_0 -1}$, i.e. we do not distinguish an extra space and the observable 3-dim space of our Universe. In fact, the huge difference in their sizes has to be explained.
As was shown in \cite{BR06}, the reason of their difference is laid in initial conditions. If an initial metric satisfies the conditions
\begin{equation}\label{times2}
M_0 = M_1\times K_1 ;\quad R_M \ll R_K ,
\end{equation}
where $R_M , R_K$ are the Ricci scalar of the main space and the extra space correspondingly, their following evolution is different. Relaxation time in the extra space $K$ is much smaller so that an evolution of metric of the main space $M$ proceeds at (almost) stationary extra metric. Initial conditions dictate a shape of the extra space while the latter influences a dynamics of the main space.

\section{Explicit form of secondary parameters}

To illustrate the above, consider a toy model with the specific pure gravitational action in a $D_0-$dim space
\begin{equation}\label{act1}
S=\int d^{D_0} z\sqrt{G}F(R_{M_0},\{b\} ) ;\quad F(R,\{b\} )=\sum_{n=1}^N b_n R^n.
\end{equation}
Here $R_{M_0}$ is the Ricci scalar and $\{b\} = b_1, b_2 , ... b_N$ .

We follow only those geometries that represent a direct product
\begin{equation}\label{UMK}
U=M_4 \times M \times K
\end{equation}
and satisfy the inequality written in (\ref{times2}). A classical motion of the metric $g_K$ as a function in the extra space $M$ is of interest and we will omit any mentioning about our 3-dim space.

Metric
\begin{eqnarray}\label{interval}
ds^{2}&=&dt^2  - g_{ab}(t,y) dy^{a}dy^{b}-
e^{2\beta_1 (t)}\gamma_{1,ij}(z)dz_1 ^{i}dz_1 ^{j} - e^{2\beta_2 (t)}\gamma_{2,ij}(z)dz_2 ^{i}dz_2 ^{j}.
\end{eqnarray}
where  $g_{ab}(t,y)$ is the metric in $M$,
$\gamma_{1,2;ij}(z)$ are positively defined internal metrics of the extra space $K=K_1 \times K_2$ and $e^{2\beta_{1,2}(t)}$ are scaling factors (see \cite{Carroll,Brokoru07}) of the spaces $K_{1,2}$. Also, $D=dim M, d=dim K$.

Evolution of the extra space metric is governed by classical equations for the functions $\beta_{1,2}(t)$. As was shown in \cite{BR06}, the effective action for the field $\beta$ has the form
\begin{eqnarray}\label{E-H_ActionD-1}                                                    
   &&S = \frac{1}{2} V [d_1] \int d^{D}y\sqrt{ \left|G^{(D)}\right|}\, \{ \sign F'\cdot [R_4 + K] - 2V(\phi) \} \\
    &&K_E =\frac{1}{d} \left( \partial\sigma
            + \frac{F''}{F'}\partial\phi \right)^{\! 2}
                    + \left(\frac{F''}{F'}\right)^{\! 2} (\partial\phi)^2
         + \sum_i d_i (\partial\beta_i)^2,                                 \label{KE}
\\
     &&-2V_E (\phi_i) =                                     \label{VE}
            e^{-\sum_i \beta_i d_i} |F'|^{-d_0/d} F(\phi),
\end{eqnarray}
\begin{equation} \label{def-phi}
      \phi_{1(2)}(t) := (d_{1(2)} - 1) e^{-2\beta_{1(2)}(t)}, \quad
      \phi := d_1\phi_1 + d_2\phi_2 .
\end{equation}

As was shown in \cite{RuInfl} the lines $\phi_{1,2}^* =0$ are attractors of the classical equations for the fields $\phi_{1,2}$. We are interested in solutions of the form $\phi_{1} \rightarrow 0; 0<\phi_{2}<\infty$.

Under these assumptions the Ricci scalar 
\begin{equation}\label{ricci}
R = R_{M} +  \phi_1 (t) + \phi_2 (t)
\end{equation}
is easily obtained.
After some algebra with formulas (\ref{ricci}) and (\ref{act1}) the reduced action can be obtained
\begin{equation}\label{act2}
S=\int d^{D_M} y\sqrt{g_M}F(R_{M},\{a\} ),
\end{equation}
where the intermediate set of parameters ${a}$ is connected to a set of primary parameters ${b}$
\begin{equation}\label{newparam}
a_n (t) = \int d^{D_K}z \sqrt{g_K }\frac{F^{(n)}( \phi_1  (t) + \phi_2 (t),\{b\})}{n!}.
\end{equation}
$g_M , g_K$ are determinants of the metric tensors of spaces $M$ and $K$.
This formula may be considered as the connection between the primary set $b$ and the metric $g_K$ of extra space $K$. Variety of the sets ${b}$ can be obtained by continuous variation the metric $g_K$ at fixed intermediate parameters ${a}$. This illustrates the hypothesis:

\emph{In the framework of multidimensional gravity, the observable set of effective parameter values could be obtained by a continuous set of initial parameter values.}

Formula (\ref{newparam}) indicates that all physical parameters vary with time. Their modern values represent the asymptotes at time tends to infinity.

In this paper we discuss the idea of "inverse landscape". By this is meant that quantum fluctuations of extra space metric and their subsequent classical evolution could lead to observable values of physical parameters in wide range of initial parameter values of a primary lagrangian. In this case the fine tuning problem seems solvable.

The study was supported by The Ministry of education and science of Russian Federation, project 14.A18.21.0789.

\end{document}